# Observation of Chiral Bound States in the Continuum in Self-biased Magneto-optical Photonic Crystals

Maohua Gong[1,†], Qiutong Zhen[2,†], Yujie Tang[2], Peng Hu[3], Qing-An Tu[1], Yan Meng[4,*], Peiheng Zhou[2,†], Zhen Gao[1,‡]

[1]*State Key Laboratory of Optical Fiber and Cable Manufacturing Technology, Department of Electronic and Electrical Engineering, Guangdong Key Laboratory of Integrated Optoelectronics Intellisense, Southern University of Science and Technology, Shenzhen 518055, China*

[2] *National Engineering Research Center of Electromagnetic Radiation Control Materials, State Key Laboratory of Electronic Thin Film and Integrated Devices, University of Electronic Science and Technology of China, Chengdu 610054, China*

[3]*School of Science, Chongqing University of Technology, Chongqing 400054, China*

[4] *Marine Science and Technology Domain, Beijing Institute of Technology, Zhuhai 519088, China*

[†]*These authors contributed equally: Maohua Gong and Qiutong Zhen.*

Chiral bound states in the continuum (BICs) are confined photonic modes with infinite quality factors and chiral response, offering significant potential for chiral optics. Although a novel type of spin-orbital-locking chiral BIC was recently predicted in magneto-optical (MO) photonic crystals (PhCs) that break time-reversal symmetry (TRS), its experimental realization has remained elusive. Here, we report the first experimental observation of such chiral BICs in self-biased MO PhCs, which operate without external magnetic fields. Moreover, we experimentally demonstrate that the chirality and the surrounding near-circular polarization of these chiral BICs can be switched simply by reversing the remanent magnetization, without any structural changes. Unlike conventional chiral BICs that preserve TRS, these magnetically induced chiral BICs exhibit exceptional robustness against structural imperfections and perturbations. This work represents a significant advancement in the topological photonics of chiral BICs, opening new pathways toward robust chiral optical devices.

The pursuit of perfect light confinement is a central goal in optics, driven by applications ranging from lasers and optical modulators to sensors and quantum technologies [1-5]. Conventional approaches, such as total internal reflection and photonic band gaps, achieve confinement but are limited by the finite lifetimes of photons, which restrict their scalability and efficiency [6,7]. In contrast, bound states in the continuum (BICs)—nonradiative modes that coexist with the radiation spectrum—can achieve theoretically infinite quality ($Q$) factors by decoupling from all radiative channels [8-13]. This unique combination of strong light localization and topological robustness has spurred widespread research, enabling breakthroughs in nonlinear harmonic generation [14,15], high-$Q$ lasers [16-18], sensing [19-21], quantum light sources [22,23], and topological photonics [24-26]. Over the past decade, diverse BICs—including symmetry-protected, Friedrich–Wintgen, and topologically protected— have been experimentally realized, firmly establishing them as a versatile platform for light trapping and wave control in micro- and nano-photonics [27-34].

Going beyond conventional BICs, chiral BICs introduce an additional degree of freedom by merging perfect field confinement with asymmetric coupling to circular polarizations [35-44]. This inherent chirality fundamentally distinguishes them from ordinary BICs and enables novel functionalities such as spin–orbit–locked emission [35,37], circularly polarized lasing [45,46], robust $C$-point generation [40], and intrinsic chiral light–matter interactions [42-44]. Conventional realizations of chiral BICs typically rely on simultaneously breaking in-plane and out-of-plane symmetries while preserving time-reversal symmetry (TRS). However, such an approach often yields chiral BICs that exhibit limited $Q$ factors and are highly sensitive to structural imperfections [41,43]. Recent theoretical breakthroughs have proposed a new solution: breaking TRS with an external magnetic field lifts the doubly degenerate symmetry-protected BICs into a pair of spin-orbital-locking chiral BICs with opposite chirality and ultrahigh $Q$ factors [35,37]. However, to date, despite these theoretical predictions, experimental observation of these magnetically induced chiral BICs has remained elusive.

In this Letter, we report the first experimental observation of a pair of robust chiral BICs with opposite chirality in self-biased magneto-optical (MO) photonic crystals (PhCs), requiring no external magnetic field. The modes near these chiral BICs exhibit near-circular polarization. Furthermore, we demonstrate that reversing the remanent magnetization alone switches the chirality of the BICs and their surrounding circularly polarized modes, without modifying the photonic structure. Unlike conventional chiral BICs with preserved TRS, our magnetically induced chiral BICs originate from a fundamentally different mechanism with broken TRS. The breaking of TRS grants them intrinsic chirality, topological robustness, and tunability, establishing a new paradigm for chiral emitters and nonreciprocal topological photonic devices.

We begin with a hexagonal two-dimensional (2D) MO PhC illustrated in the upper panel of Fig. 1(a). In the unmagnetized state ($M_r = 0$), the system possesses $C_{6v}$ and TRS, resulting in a pair of degenerate symmetry-protected BICs at the Γ point (grey dot, middle panel). The far-field polarization of the lower band (lower panel) is predominantly linear near the Γ point, featuring a polarization singularity characteristic of a BIC. When a positive remanent magnetization ($M_r > 0$) is applied (white arrows, upper panel of Fig. 1(b)), the TRS is broken while the $C_{6v}$ symmetry is maintained. This lifts the degeneracy at the Γ point and splits the degenerate BIC into a pair of chiral BICs (red and blue dots in the middle panel of Fig. 1(b)) with opposite chirality (see Supplemental Note 1 for symmetry analysis). Consequently, the far-field polarization textures near the Γ point for the lower and upper bands now exhibit pseudospin patterns corresponding to left-handed (LCP, blue circles) and right-handed (RCP, red circles) circular polarizations, respectively, as shown for the lower band in the lower panel of Fig. 1(b).

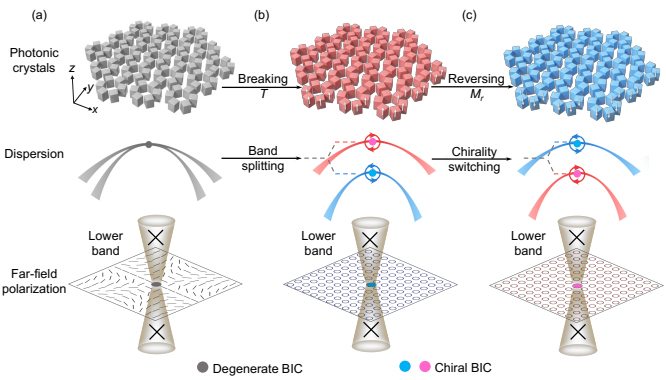

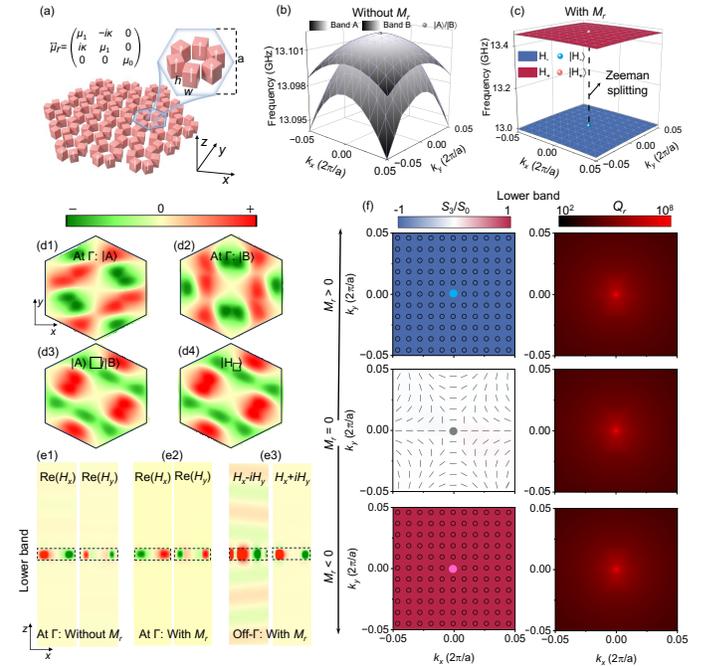

FIG. 1. Physical mechanism of the magnetically induced chiral BICs with TRS breaking. (a) A 2D MO PhC (upper panel) with $C_{6v}$ and TRS hosts a pair of doubly degenerate symmetry-protected BICs (grey dot in the middle panel) at the Γ point. The far-field linear polarization of the lower band (lower panel) exhibits an integer topological charge around the Γ point, verifying the existence of a pair of degenerate symmetry-protected BICs. (b) When breaking TRS by a non-zero remanent magnetization along the $+z$ direction ($M_r > 0$, white arrows in the upper panel), the doubly degenerate BICs split into a pair of chiral BICs (red and blue dots in the middle panel) with opposite chirality. The far-field polarization near the Γ point evolves from linear to nearly circular (bottom panel). (c) The chirality of chiral BICs and their surrounding far-field polarization can be switched by reversing the magnetization direction (white arrows) without modifying the PhC structures.

Interestingly, reversing the direction of the remanent magnetization (white arrows, upper panel of Fig. 1(c)) switches the chirality between the upper and lower bands (middle and lower panels), thereby enabling in-situ control of both the chiral BICs and their surrounding near-circularly polarized modes without requiring structural alterations. Furthermore, the demonstrated chiral BICs exhibit significantly enhanced robustness originating from TRS breaking—a mechanism fundamentally distinct from the conventional paradigm of creating chiral BICs by shifting polarization singularities (C-points) toward the Γ point [43]. Figure 2(a) illustrates the schematic of the MO PhC with a positive remanent magnetization ($M_r > 0$). The unit cell (blue frame) has a lattice constant of $a$ = 15 mm and consists of six identical MO pillars, each with a height of $h$ = 6 mm and a width of $w$ = 3.4 mm. To facilitate experimental validation, we employ lanthanum-doped barium hexaferrite (La-BaM) to break the TRS, a self-biased MO material with a permittivity of $\varepsilon = 23.2$ and a high remanent magnetization $M_r$ oriented along the $z$ direction, which gives rise to pronounced off-diagonal components in its permeability tensor [35–37, 47]:

FIG. 2. Numerical simulation of the chiral BICs in self-biased MO PhCs. (a) Geometrical parameters and permeability of the self-biased MO PhC. The inset shows the unit cell, with the white arrows indicating the direction of remanent magnetization $M_r$. (b) Simulated band structure for $M_r = 0$, where two eigenmodes degenerate at the Γ point (grey dot). (c) Simulated band structure for $M_r \neq 0$, the two degenerate eigenmodes at the Γ point split into a pair of chiral BICs (red and blue dots). (d) Eigenmode profiles of the two degenerate modes $|A\rangle$ (d1) and $|B\rangle$ (d2) at the Γ point under $M_r = 0$. The superposed field distribution $|A\rangle \pm i|B\rangle$ in (d3) corresponds to the chiral eigenmode profiles $|H_\pm\rangle$ obtained for $M_r > 0$ in (d4). (e1-e2) Side views of the magnetic field distributions ($H_x$ and $H_y$) for the BIC eigenmodes at the Γ points of the lower band under $M_r = 0$ (e1) and $M_r > 0$ (e2). (e3) Side views of the circularly polarized components ($H_\pm$) of the leaky modes at the off-Γ point, exhibiting chiral selectivity under $M_r > 0$. (f) The first column shows the far-field polarization distributions near the Γ point of the lower band under $M_r = 0$ (middle row), $M_r > 0$ (top row), and $M_r < 0$ (bottom row), respectively. The background color denotes the normalized third Stokes parameter ($S_3/S_0$), characterizing the ellipticity and chirality of the modes. The second column presents the radiation $Q$ factor, showing divergent values at the Γ point for both degenerate (middle panel) and chiral BICs (upper and lower panels).

$$\overleftrightarrow{\mu}_r = \begin{pmatrix} \mu_1 & -i\kappa & 0 \\ i\kappa & \mu_1 & 0 \\ 0 & 0 & \mu_0 \end{pmatrix}, \quad (1)$$

where $\kappa$ characterizes the strength of MO coupling, $\mu_1$ is the in-plane ($x$-$y$) permeability (see Supplemental Note 2 for the measured $\kappa$ and $\mu_1$), and $\mu_0 = 1$. Fig. 2(b) shows the simulated 3D photonic band structure of the MO PhC with $\kappa = 0$ (corresponding to $M_r = 0$) via a finite-element method (COMSOL Multiphysics). The intersection point (indicated by a grey dot) of the two gray surface bands corresponds to two degenerate eigenmodes $|A\rangle$ and $|B\rangle$ at the Γ point, their $E_z$ field distributions at 13.213 GHz are presented in Figs. 2(d1) and 2(d2), respectively. These two degenerate eigenmodes correspond to high order $d_{xy}$ and $d_{x^2-y^2}$ orbits protected by $C_{6v}$ symmetry [48]. Moreover, the far-field distributions ($H_x$ and $H_y$) of the eigenmode of the lower band are shown in Fig. 2(e1), both of which exhibit strong confinement within the MO pillars (dashed rectangle) and negligible radiation into free space, demonstrating the bound nature of the eigenmode.

In addition, BICs can be also identified through singularities in the far-field polarization states, with the corresponding topological charge defined as [49-51]:

$$q = \frac{1}{2\pi} \oint_L d\boldsymbol{k}_\| \nabla_{\boldsymbol{k}_\|} \varphi(\boldsymbol{k}_\|) \quad (2)$$

Where $L$ is a closed loop encircling the singularity in momentum space, and $\varphi(\boldsymbol{k}_\|)$ denotes the azimuthal angle of the major axis of the polarization ellipse. The far-field polarization distributions of the lower band of the MO PhC with $M_r = 0$ are shown in the middle panel (left column) of Fig. 2(f). A polarization singularity (grey dot), surrounded by nearly linearly polarized states, is clearly observed and carries a topological charge of $q = -2$. Together with the ultrahigh radiation $Q$ factor, as shown in the middle right of the right column, confirms the existence of a pair of doubly degenerate symmetry-protected BICs at the $\Gamma$ point (see Supplemental Note 3 for the corresponding simulation results of the upper band).

When the TRS of the MO PhC is broken by a nonzero $\kappa$ in Eq. (1), corresponding to a positive magnetization ($M_r > 0$) along the $+z$ direction, a Zeeman-like splitting of the originally doubly degenerate BICs emerges in the simulated 3D photonic band structure, as shown in Fig. 2(c). In this scenario, the two originally degenerate eigenmodes $|A\rangle$ and $|B\rangle$ become coupled, resulting in two hybridized modes $|H_\pm\rangle$, which can be decomposed into two orthogonal circularly polarized components:

$$H_\pm = \langle \hat{e}_x \pm i\hat{e}_y | \boldsymbol{H} \rangle, \quad (3)$$

where $\hat{e}_x$ and $\hat{e}_y$ are unit vectors along the $x$ and $y$ directions, respectively, and $H_+$ ($H_-$) corresponds to RCP (LCP) in the $xy$ plane. The two hybridized modes with superposed profiles $|A\rangle \pm i|B\rangle$ [Fig. 2(d3)] show excellent agreement with the simulated coupled eigenmodes $|H_\pm\rangle$ [Fig. 2(d4)], indicating that $|H_\pm\rangle$ correspond to circularly polarized states with opposite chirality, which can be expressed as $[1, \pm i]^T$ in the basis of $|A\rangle$ and $|B\rangle$. The far-field emission profiles ($H_x$ and $H_y$) at the $\Gamma$ point for the lower band exhibit strong confinement [Fig. 2(e2)], demonstrating the nonradiative nature of the chiral BICs. On the other hand, as shown in Fig. 2(e3), the far-field emission profiles of eigenmodes slightly away from the $\Gamma$ point, such as at wavevector $(k_x, k_y) = (0.01, 0) \times 2\pi/a$, exhibiting clear chiral selectivity: strong confinement for the RCP component ($H_+$, right panel), while evident radiation for the LCP component ($H_-$, left panel).

Besides, the emergence of magnetically induced chiral BICs is further supported by the far-field polarization singularity (blue dot) carrying a topological charge of $q = -2$ surrounded by nearly circular polarization states, and by the divergent radiative $Q$ factor at the $\Gamma$ point, as shown in the upper panel of Fig. 2(f). The chirality of the chiral BICs is indicated by the background color extracted from the normalized third Stokes parameter $S_3/S_0$, consistent with the LCP field patterns in Figs. 2(d) and 2(e) (see Supplemental Note 3 for the analysis of the upper band, which exhibits opposite chirality). Therefore, both the near-field eigenmode analysis and the far-field polarization states demonstrate that breaking TRS splits the originally doubly degenerate BICs into a pair of chiral BICs with opposite chirality: the upper band supports a right-handed chiral BIC (red dot), while the lower band supports a left-handed chiral BIC (blue dot).

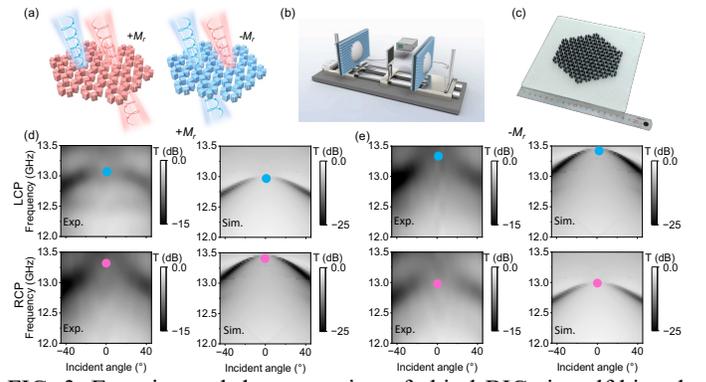

FIG. 3. Experimental demonstration of chiral BICs in self-biased MO PhCs. (a) Schematic of the MO PhC with $M_r > 0$ (left panel) and $M_r < 0$ (right panel) under oblique LCP and RCP excitations. (b) Schematic of the experimental setup consists of a vector network analyzer, an experimental sample, and a pair of circularly polarized lens antennas. (c) Photograph of the fabricated self-biased MO PhC sample. (d) Measured (left column) and simulated (right column) transmission spectra for $M_r > 0$ under oblique LCP (upper panel) and RCP (lower panel) incidences. Red and blue dots at the $\Gamma$ point indicate the chiral BICs at the upper and lower bands, respectively. (e) Measured (left column) and simulated (right column) transmission spectra under oblique LCP (upper panel) and RCP (lower panel) incidences for $M_r < 0$, exhibiting reversed chiral selectivity.

More interestingly, the far-field polarization states in momentum space can be significantly modulated by the remanent magnetization $M_r$ of the MO PhC (see Supplemental Note 4 for the detailed evolution of the polarization states with different $\kappa$). As shown in the lower panels of Fig. 2(f), reserving the remanent magnetization $M_r$ from positive to negative preserves both the topological charge of the polarization singularity and the divergent $Q$ factor at the $\Gamma$ point, but simultaneously inverting the chirality of the chiral BICs (red dot) and the surrounding near circularly polarized modes (see Supplemental Note 5 for the explanation from the aspect of effective Hamiltonian).

We then experimentally demonstrate the magnetically induced chiral BICs in a self-biased MO PhC. Figure 3(a) presents the selective chiral response of a self-biased MO PhC with broken TRS under LCP and RCP oblique incidence for positive remanent magnetization ($+M_r$, left panel). This chiral response reverses when the magnetization orientation is flipped ($-M_r$, right panel). Figure 3(b) shows a schematic of the experimental setup, which consists of a vector network analyzer (Keysight E5080) and a pair of circularly polarized lens antennas (see Supplemental Note 6 for details) to generate incident waves and measure the transmission spectra through the sample at varying incident angles. A photograph of the fabricated self-biased MO PhC mounted on a foam substrate (relative permittivity $\approx 1.05 - 0.0005i$, thickness = 5 mm) is shown in Fig. 3(c). The measured transmission spectra of the MO PhC as a function of the incident angles with positive remanent magnetization ($M_r >$

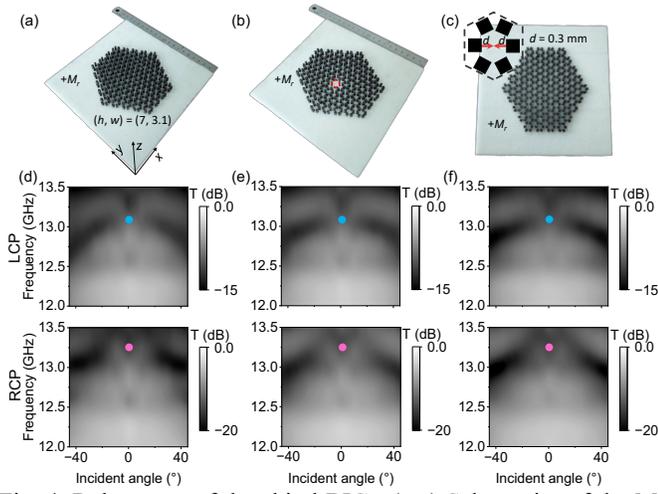

Fig. 4. Robustness of the chiral BICs. (a-c) Schematics of the MO PhC samples with different imperfections: (a) fabrication deviations with $(h', w') = (h + 1 \text{ mm}, w - 0.3 \text{ mm})$; (b) a defect induced by removing an entire unit cell (highlighted by the red dashed frame); (c) a perturbation that breaking the $C_{6v}$ symmetry with a relative displacement $d$ along the $x$ direction (indicated by red arrows in the inset). (d-f) Measured transmission spectra as a function of incident angle under LCP incidences (upper panels, blue dots) and RCP (lower panels, red dots), corresponding to the structures in (a)-(c), respectively.

0) under LCP (top row) and RCP (bottom row) incidences are shown in the left panel of Fig. 3(d). The grayscale maps of the measured transmission spectra reproduce the upper (lower) band under RCP (LCP) excitation, exhibiting clear chiral selectivity and agreeing well with the simulated results shown in the right panel of Fig. 3(d). The experimentally observed chiral selectivity, together with the transmission dips at the Γ point (red and blue dots), confirms the existence of Chiral BICs. The measured (left panel) and simulated (right panel) results for the MO PhC with reversed remanent magnetization ($M_r < 0$) are shown in Fig. 3(e). The opposite chiral response clearly demonstrates that the chirality of the BICs can be switched simply by reversing $M_r$ without any change in the structural configuration (see more details in Supplemental Note 7).

Finally, we experimentally test the robustness of the chiral BICs by introducing three types of structural imperfections into the MO PhCs: modifying the geometrical dimensions to $(h', w') = (h + 1 \text{ mm}, w - 0.3 \text{ mm})$, removing an entire unit cell, and shrinking the intracell distance along the $x$ direction, as illustrated in Figs. 4(a)-4(c), respectively. The measured transmission spectra of these imperfect samples under LCP (top row) and RCP (bottom row) incidences are shown in Figs. 4(d)-4(f). The persistence of a band exhibiting distinct chiral selectivity, along with the nonradiative character of the Γ-point BICs (blue and red dots), demonstrates the robustness of the chiral BICs in the TRS-broken MO PhC.

In summary, we have experimentally observed, for the first time, magnetically induced chiral BICs in a self-biased MO PhC operating without external magnetic fields. Our study reveals that in a $C_{6v}$-symmetric PhC, MO coupling between two degenerate BIC modes at the Γ point induces a Zeeman-like splitting, yielding a pair of chiral BICs with opposite chirality. These chiral BICs are surrounded by states with nearly circularly polarized modes. Furthermore, we experimentally demonstrate that reversing the remanent magnetization, $M_r$, switches the chirality of both the chiral BICs and the surrounding circularly polarized states. More interestingly, we experimentally verify the robustness of these magnetically induced chiral BICs against various structural imperfections—a notable advantage over conventional symmetry-protected BICs and TRS-preserved chiral BICs. These results establish a new approach to robust chiral wavefront control and are expected to advance applications across chiral optics and integrated photonics.

Z.G. acknowledges funding from the National Key R&D Program of China (grant no. 2025YFA1412300), National Natural Science Foundation of China (grants no. 62361166627 and 62375118), Guangdong Basic and Applied Basic Research Foundation (grant no. 2024A1515012770), Shenzhen Science and Technology Innovation Commission (grants no. 202308073000209), and High-level Special Funds (grant no. G03034K004). P.Z. acknowledges the funding from the National Natural Science Foundation of China (grants no. 52425205 and 52021001). Y.M. acknowledges the support from the National Natural Science Foundation of China (grant no. 12304484) and the Guangdong Basic and Applied Basic Research Foundation (grant no. 2024A1515011371).

‡gaoz@sustech.edu.cn
†phzhou@uestc.edu.cn
*mengyan@dgut.edu.cn